\documentclass[conference, 10pt]{IEEEtran}

\usepackage{cite}
\usepackage{amsmath,amssymb,amsfonts}
\usepackage{algorithmic}
\usepackage{url}
\usepackage{graphicx}
\usepackage{color}
\usepackage{placeins}
\usepackage{float}
\usepackage{tabularx,colortbl}
\usepackage{ifthen}
\usepackage{textcomp}
\usepackage{comment}
\usepackage{siunitx}
    \DeclareSIUnit{\belmilliwatt}{Bm}
    \DeclareSIUnit{\belisotropic}{Bi}
    \DeclareSIUnit{\rpm}{rpm}

\input{title.meta.tex}

\title{Observations on the Angular Statistics of the Indoor Sub-THz Radio Channel at 158 GHz}

\author[org2]{Alper Schultze}
\author[org1]{Wilhelm Keusgen}
\author[org2]{Michael Peter}
\author[org3]{Taro Eichler}

\address[org2]{Fraunhofer Heinrich Hertz Institute, Berlin, Germany}
\address[org1]{Technische Universität Berlin, Berlin, Germany}
\address[org3]{Rohde \& Schwarz, Munich, Germany}

\begin{document}

\newmaketitle

\begin{abstract}

This paper presents selected results from a channel measurement campaign conducted in a shopping mall scenario at 158 GHz. The focus is on the statistical analysis of the collected measurement data in terms of directional channel gain. Although most power is received from the line-of-sight (LOS) direction, significant multipaths arrive from all measured azimuth directions. The median of the sorted directional channel gain can be approximated by a linear curve showing an offset of about 10 dB with respect to the LOS direction.
\end{abstract}

\section{Introduction}


Previous research related to 5G millimeter-wave systems has shown that (adaptive) narrow-beam antennas have a major impact on the radio channel---the "effective channel" consisting of transmit antenna, propagation channel and receive antenna \cite{bib1}. Recent discussions on the sixth generation of mobile networks (6G) have further increased the importance of such studies and accurate modeling approaches for the sub-THz and THz frequency range. Based on channel measurements, this paper investigates the impact of multipath propagation at \SI{158}{\giga\hertz} in a shopping mall scenario and the directional channel gain (normalized received power) that a system with beamsteering functionality would observe by exploiting the receive directions with the strongest reflection paths. 

\section{Channel Sounder Setup}

The instrument-based time-domain channel sounder makes use of D-band front ends acting as up-/down-converters from/to an intermediate frequency (IF) stage and was operated at a carrier frequency of \SI{158}{\giga\hertz} with a measurement bandwidth of \SI{2}{\giga\hertz}.
The complex sounding sequence, which has a duration of \SI{100}{\micro\second}, is provided on the IF by a wideband vector signal generator. On the receiver side, a multitude of signal periods are sampled in the IF domain using a signal analyzer. A rotation table enables angle-resolved measurements by turning the down-converter in the azimuth plane. The down-converter was equipped with a standard gain horn antenna with a gain of \SI{20}{\deci\belisotropic}. The channel sounder setup's principle and the evaluation of another indoor measurement campaign are discussed in \cite{bib2} in more detail.

\section{Measurement Scenario and Procedure}

\begin{figure}[htbp]
\centerline{\includegraphics[width=0.8\linewidth]{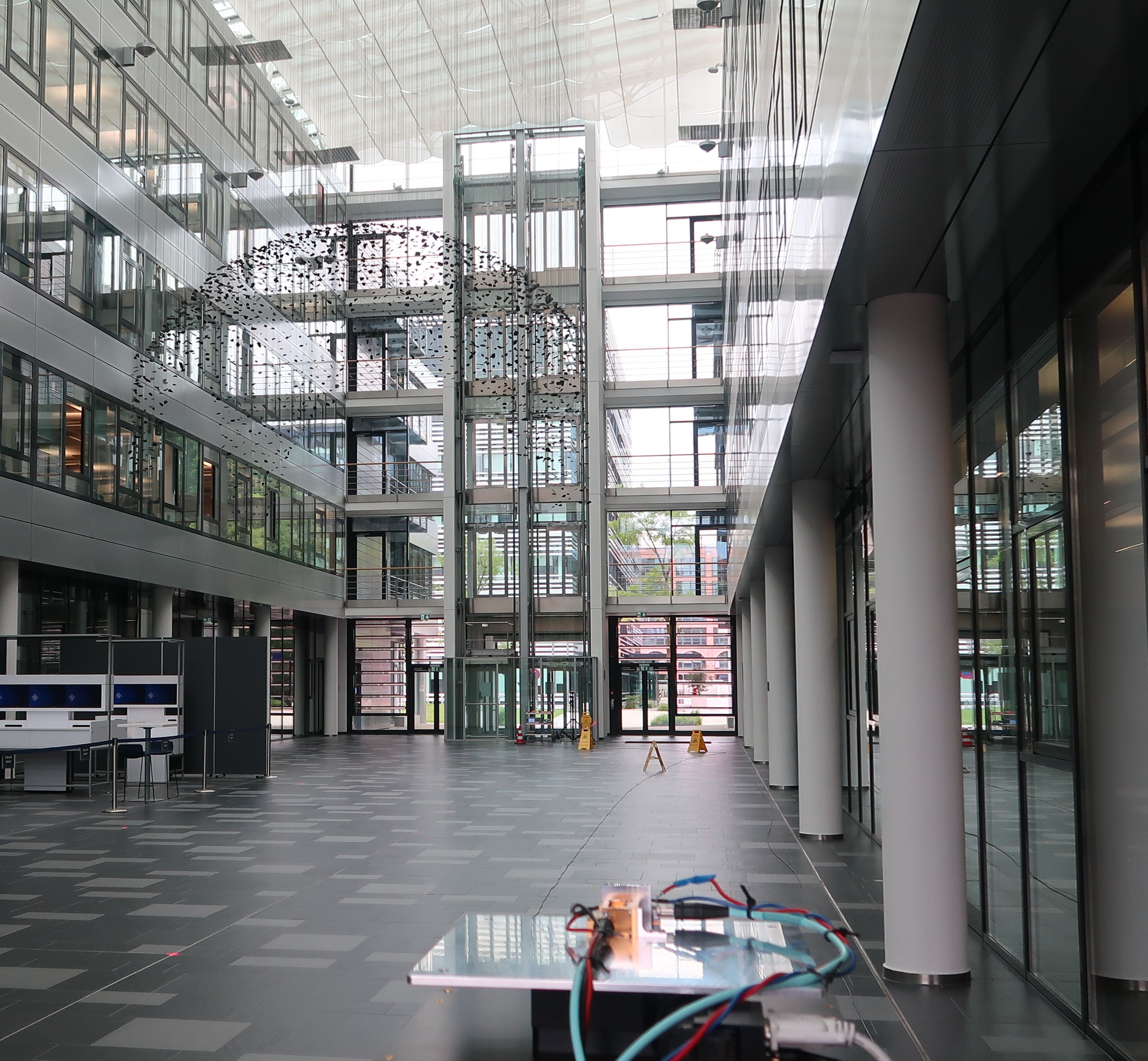}}
\caption{Shopping mall measurement scenario showing the receiver setup in front and transmitter setup in back.}
\label{fig:photo}
\end{figure}

The measurement's venue was a company building's atrium that well represents a shopping mall scenario. The room size is \SI{15}{\meter} x \SI{50}{\meter} with a ceiling height of \SI{20}{\meter}. Continuous glass fronts and a tiled floor characterize the room's procurement. The transmitter was placed centrally at the beginning of the atrium. The receiver was moved along a well defined grid of equally distributed measurement positions. A total number of $21$ measurement positions, each comprising of $24$ angles (\SI{-180}{\degree} to \SI{180}{\degree} in \SI{15}{\degree} steps), were measured. Fig. \ref{fig:photo} visualizes the shopping mall measurement scenario.

\section{Measurement Evaluation and Results}

\begin{figure}[htbp]
\begin{center}
  \includegraphics[scale=1]{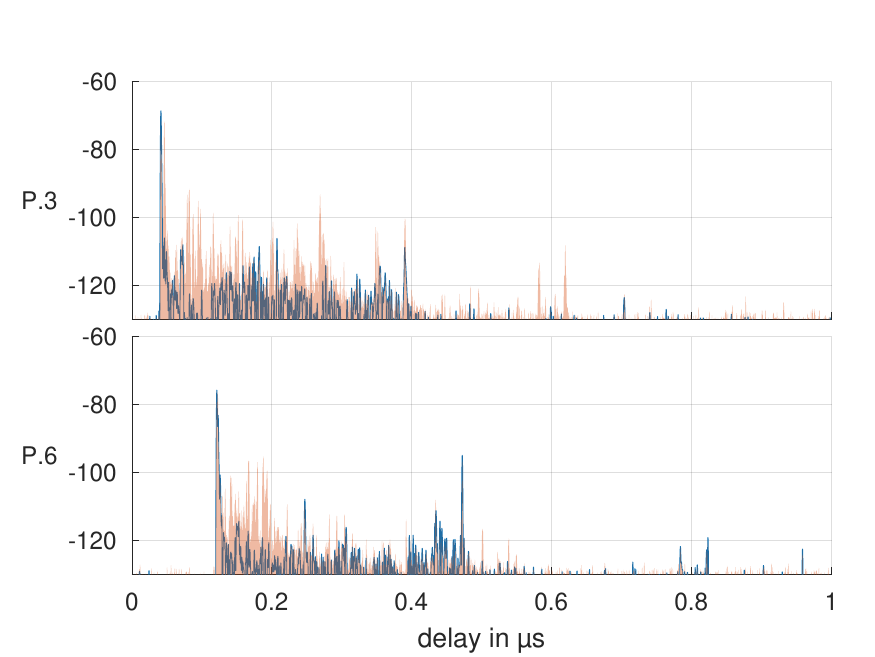}
  \setlength{\abovecaptionskip}{-0.4cm}
  \caption{CIR of LOS direction (blue) and pseudo-omnidirectional CIR (red) for P.3 (\SI{12.4}{\meter} LOS distance) and P.6 (\SI{36.3}{\meter} LOS distance). The CIRs are plotted as magnitude in dB over delay in µs.}
  \label{fig:cir_p3_p6}
\end{center}
\end{figure}

\begin{figure}[htbp]
\begin{center}
  \includegraphics[scale=1]{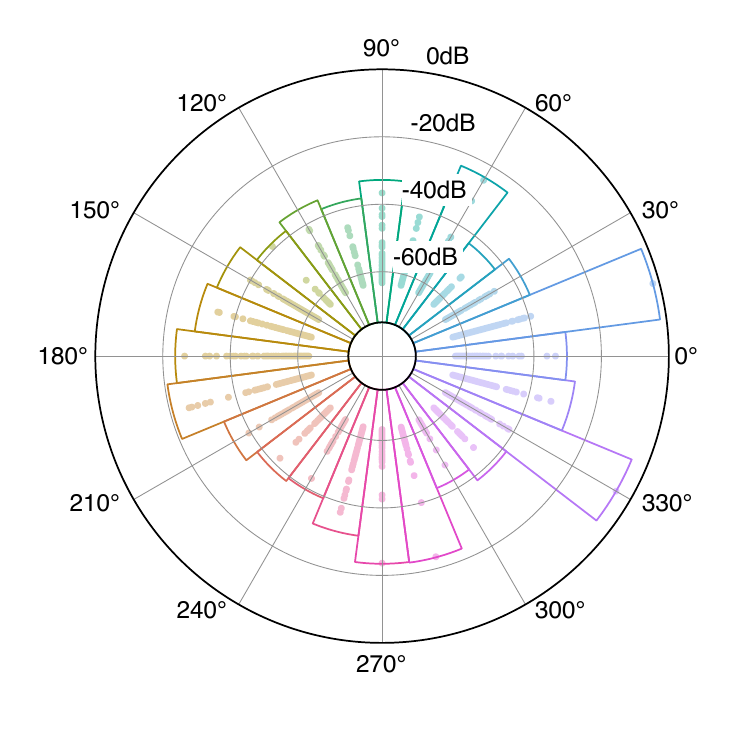}
  \setlength{\abovecaptionskip}{-0.5cm}
  \caption{Rose plot of normalized directional channel gains for measurement position P.3. The individual path gains are symbolized as dots. Normalization has been done with respect to 0 dB omnidirectional channel gain.}
  \label{fig:rose_p3}
\end{center}
\end{figure}

\begin{figure}[htbp]
\begin{center}
   \includegraphics[scale=1]{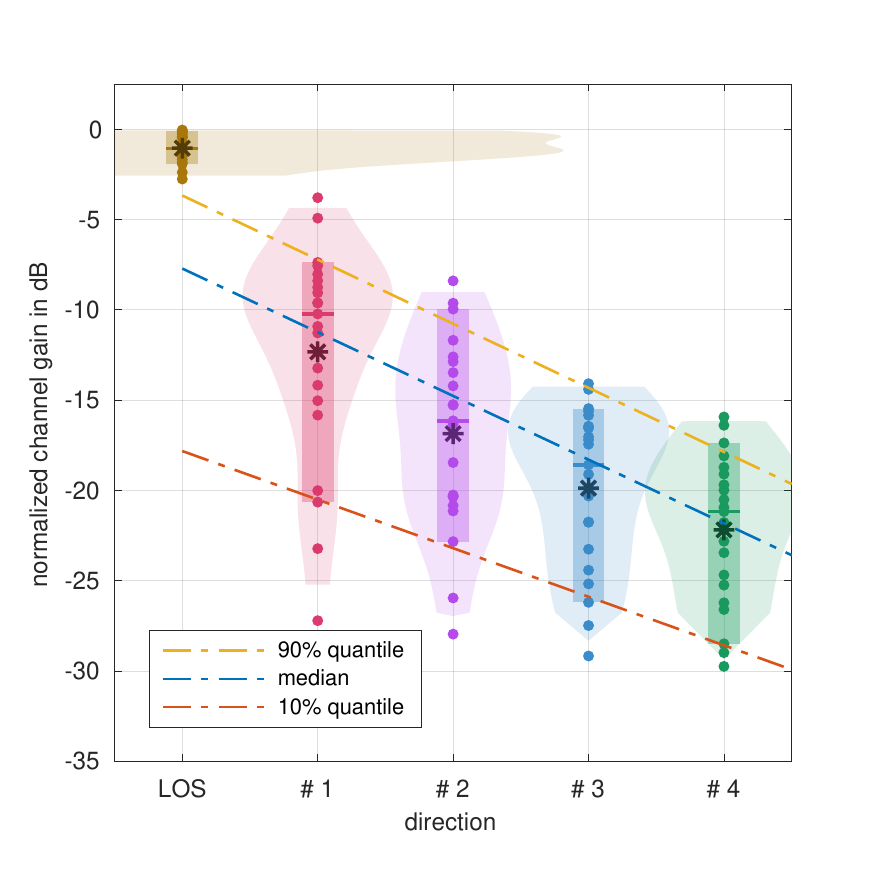}
   \setlength{\abovecaptionskip}{-0.7cm}
   \caption{Violin plot of the five strongest directions including LOS direction. Dots indicate the sampled normalized channel gains per measurement position, stars the mean values. The bar and the bold line show the 10\% ($10^\mathrm{th}$ percentile), 50\% (median), and 90\% quantiles ($90^\mathrm{th}$ percentile). The dotted lines are the linear fits of the quantiles.}
   \label{fig:violin}
\end{center}
\end{figure}

In the first evaluation step, the received sounding sequences were subjected to a phase compensation, coherent averaging, and correlation with a pre-recorded back-to-back calibration, resulting in one channel impulse response (CIR) per measurement position and direction (see Fig. \ref{fig:cir_p3_p6}).

In the second step, a discrete path estimation in the angular-delay domain was performed by means of a local peak search \cite{bib2}, leading to complex path coefficients being discretized into (cyclic) angular bins with \SI{15}{\degree} width and delay bins with \SI{0.5}{\nano\second} width. The  estimation was performed with respect to a fixed threshold of \SI{-130}{\decibel}, which is significantly higher than the noise floor (compare Fig. \ref{fig:cir_p3_p6}).
The sum of path gains in one direction yields the directional channel gain, whereas the sum over all directions yields the omnidirectional channel gain. In this paper, we are more interested in the underlying directional statistics, thus we apply a normalization to the omnidirectional channel gain in the sequel.

Fig. \ref{fig:rose_p3} shows estimated directional channel gains and path gains per angular bin for position P.3 in form of a rose plot.  

One can clearly see, that the direction \SI{15}{\degree} (LOS direction) shows the largest channel gain. However, the channel gain for \SI{330}{\degree} is only \SI{3}{\decibel} smaller in this example. Furthermore, individual path gains per direction may vary significantly.


In the following, we investigate the distribution of the directional channel gains of the strongest directions across all measurement positions. 
Fig. \ref{fig:violin} shows the statistical evaluation in form of a violin plot \cite{bib3}  for the strongest direction, i.e. LOS direction and the following next strongest secondary directions. For all measurements the LOS path contains the highest directional channel gains. Table \ref{tab:statistics} summarizes the results for median and $10^\mathrm{th}$ and $90^\mathrm{th}$ percentile as dB values. Additionally, a linear regression across the order was performed for all secondary directions.

%
\begin{table}[htbp]
\begin{center}
\caption{Statistics of normalized channel gain in dB per direction} 
\label{tab:statistics}
\begin{tabular}{|p{2cm}|l|l|l|}
 \hline
 & median & $10^\mathrm{th}$ percentile & $90^\mathrm{th}$ percentile\\
 \hline
 LOS direction & $-1.1$ & $-1.9$ & $-0.1$\\
 \hline
 Secondary direction of order $N$ & $-3.5 N - 7.7$ & $-2.7 N - 17.8$ & $ -3.6 N -3.7.6$\\
 \hline
\end{tabular}
\end{center}
\end{table}

\section{Discussion}
The observed angular distribution of directional channel gains is important for a communication system utilizing beamforming. Assuming beamforming with 15° beamwidth, the results from Table \ref{tab:statistics} directly show the expected relative powers for LOS and the next strongest beam directions, which become relevant in case of blockage. The distribution of individual path gains within one (beam) direction (as seen in Fig. \ref{fig:rose_p3}) gives hints on the expected frequency selectivity of the beamforming channel.



\end{document}